\documentclass{llncs}
\usepackage{lncsedoc}
\usepackage[final]{graphicx}
\begin{document}
\title{Quantum objects as elementary units of causality and locality}

\author{ Hans H. Diel}

\institute{Diel Software Beratung und Entwicklung, Seestr.102, 71067 Sindelfingen, Germany,
diel@netic.de}

\maketitle

\begin{abstract}
The author's attempt to construct a local causal model of quantum theory (QT) that includes quantum field theory (QFT) resulted in the identification of “quantum objects” as the elementary units of causality and locality. Quantum objects are collections of particles (including single particles) whose collective dynamics and measurement results can only be described by the laws of QT and QFT. Local causal models of quantum objects' internal dynamics are not possible if a locality is understood as a space-point locality. Within quantum objects, state transitions may occur which instantly affect the whole quantum object. The identification of quantum objects as the elementary units of causality and locality has two primary implications for a causal model of QT and QFT: (1) quantum objects run autonomously with system-state update frequencies based on their local proper times and with either no or minimal dependency on external parameters.
 (2) The laws of physics that describe global (but relativistic) interrelationships must be translated to a causal model of interactions between 
quantum objects and interactions between quantum objects and the space.
\end{abstract}


\section{Introduction}

The subject of this article is quantum objects, and they have been introduced by the author in the course of his attempts to construct a local causal model of quantum
theory (QT) that includes quantum field theory (QFT). Quantum physicists consider Bell's famous inequality
 (see \cite{Bell}) and its violation in experiments to be a
strong indication that local causal models of QT are not possible. Based on the formal definition of a local causal model, the author came to the conclusion that
local causal models of QT/QFT are indeed impossible if the notion of a locality is understood as a space-point locality   (see \cite{DielLC}). 
However, if certain QT deficiencies are removed, causal models of QT/QFT where the
non-localities are confined to elementary units called quantum objects, appear to be feasible. In the present article,
quantum objects are discussed in more detail.

Quantum objects were introduced in
 \cite{DielLC} as part of a causal model of QT/QFT. Therefore, the article starts with a description of causal models in
general and of the proposed causal model of QT/QFT, which surrounds quantum objects. Section 2 mainly summarizes material that is described in more detail in 
 \cite{DielLC}. 
As described in \cite{DielLC} (and summarized in Section 2), a formal local causal model of an area of physics has 
three types of constituents: (1) the physics engine,
(2) the laws of physics, and (3) the system state referred to by the laws of physics. The composition of these three components determines whether
a local causal model is feasible, or more specifically, which type of locality and causality can be achieved. In
  \cite{DielLC}, it has already been concluded that a local
causal model of QT/QFT is not possible if a locality is understood as a space-point locality. Thus, a local causal model of QT/QFT is not possible if
the system state consists solely of space points and their associated attributes. In addition, the (causal) laws of QT/QFT need to refer to aggregate objects and parameters, such as particles, waves, and fields; moreover, the system state needs to contain these objects and parameters. Quantum objects have been introduced as a generalization of the types of objects that are known in standard
QT.
The assumption that quantum objects are autonomous and depend as little as possible on the external system state parameters enables the confinement of nonlocal
causal state progressions to quantum objects' internal processes. 

The tolerance of non-localities within quantum objects results in deviations
from relativity theory within quantum objects. This deviance has implications for the objects' internal space-time concept that are addressed in Section 6.
Although the focus of this article is on quantum objects' internal processes, it is also necessary to consider the global interrelationships among quantum
objects. These interrelationships are primarily determined by the interactions between quantum objects and the interactions between quantum objects and fields. This subject is
addressed in Section 4.

The causal model of quantum objects may be viewed as another representation of QT/QFT with a focus on specific aspects of QT. Ideally, another
view or model would not provide any new knowledge if the underlying theory (i.e., QT) were properly and completely defined in all areas. However, because
of certain QT/QFT deficiencies  (see Section 2.2), the proposed causal model of QT/QFT necessitates specifications that cannot be derived from standard QT/QFT. In part,
the proposed causal model has been developed with the goal of removing some of the known QT deficiencies. Similarly, other QT/QFT limitations have been
detected in the course of the model's development, and solutions for the removal of these limitations have been included.

\section{Causal Models}

In \cite{DielLC},  a causal model of an area of physics is defined as consisting of three types of constituents:
\begin{enumerate}
\item The "physics engine" that defines the overall (i.e., subject-independent) interpretation of the (subject-dependent) laws of physics;
\item The laws of physics that describe for the subject area of physics how the state of the system evolves under various circumstances; and
\item The system state, i.e., the objects and elements that are referenced by the laws of physics.
\end{enumerate}
\begin{figure}[ht]
\center{\includegraphics*[scale=0.5] {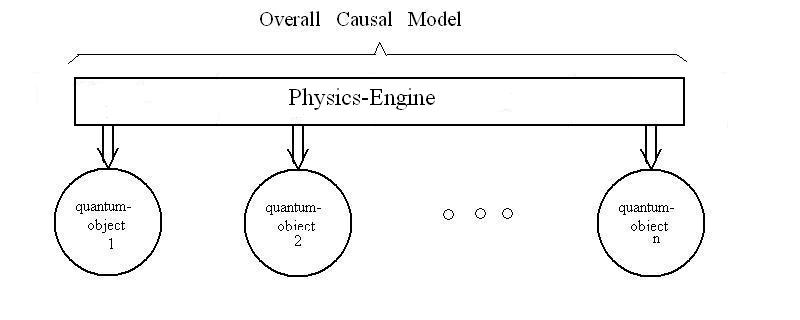} }
\caption{Causal model with physics engine performing total system state updates}
\end{figure}
The physics engine acts upon the state of the physical system and continuously determines new states in uniform time steps. For the formal
definition of a causal model of a physical theory, the continuous repeated invocation of the physics engine to realize the progression of the state of the system
is assumed.
\\
\\
$ systemEvolution( system\; \; S ) := \{    \\
 S.t = 0; S.x_{1}=0; S.x_{2}=0; S.x_{3}=0;  \\
 S.\psi = initialState; \\
 \Delta t = timestep;     ||\;  must\; be\; positiv  \\
 DO \;  UNTIL ( nonContinueState( S ) )  \{  $

   physics engine $(  S, \Delta t );  $ \\
\}  
\\
\}  
\\
\\
physics engine $(  S,  \Delta t ) := \{  $

$ \; \;    S = applyLawsOfPhysics( S, \Delta t );  $  \\
\}  
\\
\\
The refinement of the statement $  S = applyLawsOfPhysics( S, \Delta t )  $ provides the laws of physics for the given subject area of physics; it consists of a list of laws
 $  L_{1}, ... ,  L_{n} $
that define how an "in" state s evolves into an "out" state s.

    $      L_{1} : \: IF \: c_{1}(s)\: THEN \: s = f_{1}(s); $

    $      L_{2} : \: IF \:c_{2}(s) \: THEN  \: s = f_{2}(s); $

            ...

      $     L_{n} : \: IF \:c_{n}(s) \: THEN   \:s = f_{n}(s); $
\\
The system state specifies in more detail the objects that are referenced by the
laws of physics, their internal structure and their interrelationships. In the simplest
case, the system state consists of a set of space points with their associated contents $  \psi $.
\\
\\
$ systemstate := \{ spacetimepoint ... \} \\
\; \;   spacetimepoint :=\{  t,  x_{1}, x_{2}, x_{3}, \psi \} \\
\; \;  \psi :=  \{ stateParameter_{1}, ... , stateParameter_{n} \} $

\subsection{Local Causal models}

A causal model of a theory of physics is called a spatial causal model if (1) the system state contains a component that represents a space, and (2) all of the
other components of the system state can be mapped to the space. There exist numerous textbooks on physics (mostly in the context of Relativity theory) and
on mathematics that define the essential features of a "space". For the purpose of this article and the subject locality, it is sufficient to require that the space
(which is assumed to be part of the model) supports the notions of position, coordinates, distance, and neighborhood. The definition of a local causal model presupposes
a spatially causal model. 

A causal model is understood to be a local model if changes in the state of the system depend entirely on the local state and only affect 
the local state. The local state's changes can propagate to neighboring locations. The propagation of the state changes can reach distant locations; however, these changes must always be
accomplished through a series of state changes to neighboring locations.

Based on the formal definition of “causal model”, a formal definition of “locality” can be given. We are given a physical theory and a related spatially causal
model with position coordinates x and position neighborhood dx 
 (or $ x  \pm \Delta x $ in the case of discrete space points).

A causal model is called a “local causal model” if each of the laws $ L_{i} $ applies to no more than a single position x and/or to the neighborhood of this position  $ x  \pm dx $.
The position reference can be explicit or implicit with respect to a state component that has a well-defined position in space. Whereas references to the complete
space of a spatially extended object are considered as violations of the notion of a locality, references to specific properties of spatially extended objects do not violate this notion.

In this article, the above definition of a locality, which strictly refers to space points
 (i.e., x and   $ x  \pm dx $) is called a "space-point locality". To enable
the construction of causal models where a space-point locality is not achievable but the environment should not be classified as completely non-local, a weaker form of locality,
which we call “object-locality”, is defined. A causal model is called an "object-local" causal model if it is a space-point-local causal model, with the exception of object-internal
processes where the causal laws may refer to complete objects or sub-objects. For a given causal model, "object" has to be replaced with a specific type of
object belonging to the system state. For example, the causal model of QT/QFT (see below) is called a "quantum-object-local causal model" because the non-space-
point localities are confined to quantum objects.

\subsection{Problems Impeding Causal Models of QT/QFT}

The present article is a result of the author's attempt to analyze in more detail the problems that prevent the construction of local causal models of QT/QFT; furthermore, the author intends to overcome these problems. It is reasonable to first consider the feasibility of a causal model of QT/QFT, and in the next step
(if the first step was successful), we analyze the feasibility of a local (causal) model.
 The primary problem with a causal model of QT/QFT is not the peculiarity of QT/QFT, but the state of the
theory in certain areas. Four main "problem areas" have been identified, as shown in Table 1. The problem areas (the more detailed problems and proposed
solutions) have been described in several papers by the author 
 (see \cite{DielComp} and \cite{DielLC}). The present section and Table 1 give a summary of the findings. The four problem areas with respect to causal models also represent the major problem areas with respect to a local (causal) model.
This set of issues is summarized in column "Locality Problems" in Table 1.

\begin{table}
\caption{\label{label}Summary of the problem areas that impede a Local Causal Model of QT/QFT}
\begin{tabular} { | c | c | c | c |   }
\hline
Problem areas & Experiments  & Problems impeding & Locality       \\
                       &                       & a Causal  Model  &     Problems        \\
\hline
\hline

Measurement,	  &  all kinds of        & lack of                              & instant reduction of    \\
Interpretation of QT    &  experiments  &  agreed-upon theory        & complete wave     \\
\hline
Interference collapse  &   Double-slit    &   ill-defined interference    & instant collapse of       \\
	                         &                        &   collapse rule  & interference      \\
\hline
QFT-Interaction          & Scatterings    &  lack of  & integrals ranging     \\
      &   & "equation of motion"  & over        \\
      &   & &   complete space     \\
\hline
Entanglement    & Scatterings & lack of  &    none if     \\
     &     & correlation model  &    common paths      \\
     &  EPR-perfect  &  ("hidden variables")  &   for both      \\
     &  correlation   &         &    particles          \\
\hline
Entanglement    &   EPR-imperfect  &  lack of   &"action at      \\
   & correlation   & "communication model"  &    a distance"      \\
\hline
\end{tabular}
\end{table}  
Some explanations for the abbreviated formulations contained in Table 1 are appropriate. More details on the subject can be found in \cite{DielComp} and \cite{DielLC}.
\begin{itemize}
\item Measurement, Interpretation of QT
\\
The measurement problem of QT must still be considered unsolved. Measurements are mandatory ingredients of all types of experiments. A number
of alternative "interpretations" of QT have been proposed as solutions to the measurement problem. 
Without mentioning that there is no generally
agreed interpretation of QT (and no solution to the measurement problem), most of the proposed interpretations are not suitable bases
for a causal model. Even if we have a causal model of the QT measurement (such as the causal model proposed by the author in 
 \cite{Dielmeas} and in Section 2.3),
we still have the "locality problem". The fact that QT measurements instantly affect the complete spatially distributed wave prevents the
construction of a space-point-local model.
\item Interference collapse
\\
The standard explanation of the double-slit experiment found in most textbooks of QT refers to conditional statements such as "if it is possible to determine ..." rather
than to the physical objects and state parameters  (see \cite{Dieldslit}).  This type of explanation makes it impossible to construct a causal model. The causal model proposed by the
author in 
\cite{Dielmeas} and in Section 4.4 associates the interference collapse (as well as the wave function collapse) with the occurrence of QFT interactions.
\item QFT-interactions 
\\
These interactions are particle interactions such as scatterings that require QFT (e.g., a scattering matrix or Feynman diagrams) in their description and result calculation.
Although QFT provides an extensive and powerful framework for the treatment of this type of interaction, the present theory does not support
the translation of this framework into a causal model.
\item Entanglement
\\
Entanglement was the original area where the impossibility of a local causal model of QT was inferred. While it appears feasible to use "hidden variables" to construct a causal model that at least supports perfect correlations, the construction of a local causal model is not 
feasible.
\end{itemize}
 
\subsection{Causal Model of QT/QFT}

The causal model of QT/QFT is formulated partly in terms of discrete system state parameters (space, time, and particle paths) 
\footnote{Initially, the assumption of discrete system state parameters was made to simplify
the description. However, in certain cases, the assumption of
discreteness has physical implications that are mandatory for the functioning of
the causal model.}
 that support the mapping
of the causal model to a cellular automaton; for example, this type of mapping was described in 
 \cite{DielCALagr}.

\subsubsection{The system state} In addition to
quantum objects, the system state of the causal model of QT/QFT includes the space and fields. Although time is not considered to be a part of the system state, it is implied by the physics engine.
\\
\\
$ systemstate := \{ $

$ space := \{ spacepoint ... \} $; 

$ quantumobjects := \{  quantumobject_{1}, ..., quantumobject_{n}     \} $;

$ otherobjects $;

$ spacepoint :=\{  x_{1}, x_{2}, x_{3},  \psi \} $; 
\\ \}
\\
\\
\subsubsection{The physics engine}The physics engine (as described in Section 2) continuously interprets the laws of physics to advance the system state. As a necessary extension to the
general causal model given in Section 2, the physics engine for the causal model of QT/QFT implements a more sophisticated process structure. In Section 2, the
physics engine is described as a single global engine that proceeds the complete system state simultaneously in unique time steps (see Fig. 1). To
confine the non-localities to quantum objects and to support relativistic, proper time intervals, multiple object-related physics engines are assumed (see Fig. 2).
In addition to the quantum objects 1-n, a further object
 "space" with an associated physics engine 0 is shown in Fig. 2. See Section 6.2 for additional details.
\\
\\
$ systemEvolution( system\; \; S ) := \{  $  \\
FOR  ( all quantum objects  $ qobj[i] $  ) \{   

     run QT/QFT- physics engine(  $ qobj[i]  );  $ \\
\}  
\\
FOR  ( all otherobjects  $ otherobj[i] $ ) \{  
 
     run ...(  $ otherobj[i]  $ );  \\
\} 
\\
\}   
\\
\\
QT/QFT-physics engine(  qobj ) := \{  

$ \; \;    S = applyLawsOfQT/QFT( S, qobj );  $  \\
\}  
\\
\\
 \begin{figure}[ht]
\center{\includegraphics*[scale=0.5] {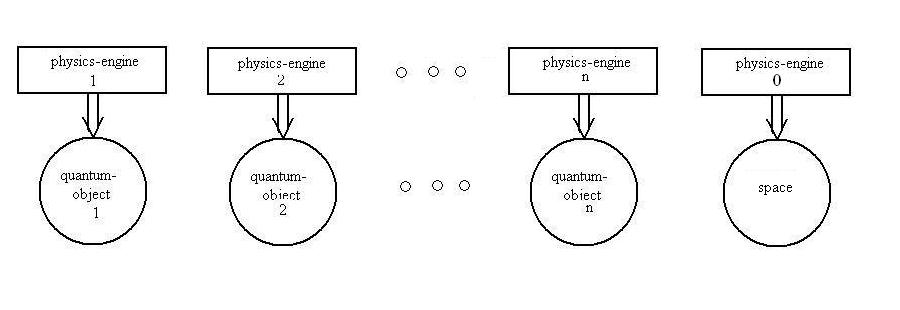} }
\caption{Causal model with a separate physics engine for each quantum object}
\end{figure}

\subsubsection{The laws of QT/QFT} The physics-engine invokes the function applyLawsOfQT/QFT() periodically with proper time intervals associated with
the respective quantum object. At the highest level of specification, applyLawsOfQT/QFT()has to determine whether the interactions between quantum
objects have to be processed first. The "normal" progression of quantum objects is deferred until all of the possible interactions have been processed.
\\
\\$ applyLawsOfQT/QFT( qobject) $  :=  \{  
 
  interaction-processing;
 
  IF ( not-destructed( qobject ) )   proceed-quantum-object( qobject );
\\
 \}
\\
For a complete causal model of QT/QFT, the functions interaction-processing and proceed-quantum-object would have to be specified in more detail. In this article, only interaction-processing is discussed in more detail (see Section 5.2).

\section{The Quantum Object}

The quantum object is the most important entity for the description of the causal model of QT/QFT. A particle may occur as a separate quantum object or 
be part of a quantum object. The following three properties distinguish quantum objects from other objects that typically occur in physics:
\begin{enumerate}
\item Quantum objects are composed of multiple alternative paths with associated probability amplitudes. With the interactions (including the measurements), the
multiple paths may be reduced to a single path.
\item Quantum objects may consist of multiple spatially separated particles.
\item Quantum objects have global attributes that apply to all of the paths and particles of the quantum object.
\end{enumerate}
The combination of these three properties make quantum objects special within physics.

\subsection{Structure, Components and State of the Quantum Object}

\begin{figure}[ht]
\center{\includegraphics*[scale=0.8] {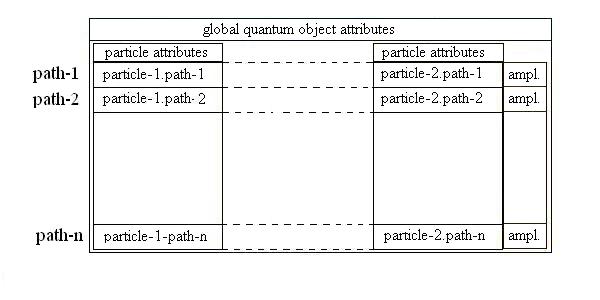} }
\caption{Structure of the quantum object consisting of two entangled particles.\label{fig1}}
\end{figure}
A quantum object may be viewed as having a two-dimensional structure. One of the dimensions represents the collection of quantum object elements, which typically consists of 1 to n particles.
\small{
\begin{verbatim}
quantum object :=   
  global-quantum-object-attributes;
  particle[1],
  ...
  particle[n];
\end{verbatim}
}
\normalsize
In the second dimension, the quantum object consists of the set of alternatives that may be selected during the evolution of the quantum object, for example,
by a measurement. In this paper, these alternatives are called "paths".
\small{
\begin{verbatim}
quantum object :=   
  global-quantum-object-attributes;
  path[1],
  ...
  path[npath];
\end{verbatim}
}
\normalsize
The two-dimensional structure is supplemented by global attributes. Whereas global quantum-object attributes are attributes that apply to the complete quantum object,
particle attributes apply to the complete particle. "Amplitude" is the single attribute that applies to a complete path. The only space-point-local attributes
are the attributes labeled "particle-i.path-j" in Fig. 3. Global attributes disturb space-point localities. Similarly, the inclusion of global attributes may be
unavoidable for the construction of causal models in theories that contain non-localities. The confinement of the non-localities of QT/QFT within
quantum objects (by assuming the global attributes) supports the view of quantum objects as the elementary units of causality and locality. Further details on
the global attributes of the quantum object are given in Section 3.3.
%
%
%

\subsection{Examples of Quantum Objects}

Different types of quantum objects can be distinguished:

\begin{itemize}
\item A single particle
\\
A single particle constitutes the simplest type of quantum object. The idea of representing a particle by a set of paths was introduced by Feynman (see \cite{Feynman1}) with the formulation of quantum electrodynamics (QED). 

\item Collections of (entangled) particles
\\
Collections of particles that can be described by a common wave function where only specific attribute combinations can occur as measurement
results represent a quantum object. Thus, the particle collection is represented by a set of paths, and each path contains the attribute combinations
for all of the particles and an associated probability amplitude (see Fig. 3). Arbitrary particle collections whose common wave function would be 
the product of the individual wave functions do not constitute quantum objects. (As a consequence, considering the whole universe
as a single large quantum object would not be in accordance with the definition of the term “quantum object” given in this paper.)
\\
The following additional types of quantum objects are special examples of particle collections.
\item Interaction object
\\
The interaction object is a type of quantum object that was described in  \cite{Dielfi} as part of a functional model of QFT interactions. It is created at the
beginning of an interaction. At the end of the interaction, the interaction object is transformed into an interaction-result quantum object.
\item Interaction result object  
\\
The result of a QFT interaction (see Section 4.2) is a quantum object containing all of the particles resulting from the interaction and the probability
amplitudes for the resulting paths. The causal model of QT/QFT assumes that the interaction object develops into the interaction result object (see Section 5.2).
\item Bound system quantum object
\\
Composite objects such as hadrons, nuclei, and atoms that are built from (elementary) particles are incorporated into the concept of a quantum object. The elements
of the composite quantum object may be grouped to form an internal structure. For example, the atom consists of the nucleus and electrons, and
the nucleus consists of hadrons. If this type of a hierarchical structure is given, only the complete (outermost) entity is called a quantum object within this
paper.

\end{itemize}
Quantum objects are dynamically created, separated and combined in specific processes such as  interactions and decays (see Sections 4 and 5).

\subsection{Object-Global Attributes and Actions}

The existence of global state information that is not merely an aggregation of local information, and therefore, cannot be broken down into finer local information
is one of the special properties of quantum objects. The quantum-object-system state described in Section 3.1 contains attributes (i.e., state parameters)
that are associated with differing entities such as the total quantum object, the particle or the path. A global attribute that is not an aggregation of space-point-local information implies that changes in the global attribute value occur instantaneously for the complete quantum object; the alternative would be the propagation
of the changes through the quantum object. Thus, global attributes are required whenever instantaneous actions that affect the complete quantum object (or the complete
particle or path) occur within the causal model. Table 2 shows the types of instantaneous actions for the four "problem areas" of the causal model of QT/QFT.

\begin{table}
\caption{\label{label}Summary of the problem areas and non-local actions}
\begin{tabular} { | c  | c | c | }
\hline
Problem areas  & instant global  & Locality scope   \\
          &     action    &     \\
\hline
\hline

Measurement,	   & instant reduction of &  single particle  \\
Interpretation of QT    & complete wave  & path set  \\
\hline
Interference collapse   & instant collapse of  &  single particle   \\
   model           & interference   & path set  \\
\hline
QFT-interaction    &   generation of   & Interaction Object  \\
        & interaction result    &  (=Quantum Object)  \\
\hline
Entanglement     &    reduction to  single &  Quantum Object  \\
with QFT-interaction     &        common path   &     \\
\hline
Entanglement      &    reduction to  single   &   Quantum Object  \\
perfect correlation     &     common path     &     \\
\hline
Entanglement     & change of  global  & Quantum Object  \\
imperfect correlation   &  angular momentum   &   \\
\hline
\end{tabular}
\end{table}  
A further reason for the inclusion of the global attributes in the system state of the quantum object is to support interactions between autonomous quantum
objects according to the laws of QFT. When quantum object 1 interacts with quantum object 2, the result of the interaction depends on quantum-object
parameters such as the energy, momentum and spin. The assumption of autonomous quantum objects implies that the respective parameters must be provided by the
interacting quantum objects rather than by the environment. This subject is elaborated on in Section 4.

For a model of an area of physics (e.g., QT), global attributes such as local (or space-point-local) attributes must have a physical representation. The definition
of “global information” implies that it is not possible to map the global attributes and parameters to space points or areas of space. Thus, the question arises of how
the global object attributes may be physically represented if we do not employ a spatial representation. Although it is not necessary to identify the physical representation of all of the parameters and attributes for the abstract formulation of the causal model of QT/QFT, it certainly increases the plausibility of the
model if at least some possible mapping of the essential state parameters can be provided. In the causal model of QT/QFT, the following possible representations
of global (i.e., space-independent) information are included:
\begin{enumerate}
\item The simplest model for the global attributes is given when the specific global attribute can be equated with an attribute that exists in classical
physics, such as the mass or angular momentum of a quantum object. However, the remaining problem is that the global attributes 
in the local causal model are typically requested to support instantaneous value changes; this feature is not supported in classical physics. Thus, it would
be a deviation from (or an extension of) QT to assume the possibility of instantaneous value changes for a specific global attribute.
\item A special type of quantum-object-global attribute is the proper time clock rate. In the causal model of QT/QFT, this attribute is assigned to the physics engine
of the quantum object. It is conceivable that additional global attributes such as the (global) momentum and angular momentum are directly assigned to
the physics engine.
\item From the point of view of QFT, there are several quantum fields associated with quantum objects. In addition to the field quanta that are associated
with the individual particles that belong to the quantum object, there could be object-global fields that span the overall quantum object, and
it may be possible to instantly change the global attributes of such fields. The respective fields may be fields known in existing QFT, for example,
the gauge field.
\end{enumerate}

\subsubsection{Global information in the Lagrangian}  Similar to most theories of physics, the descriptions of quantum theory and quantum field theories can be based on the Lagrangian(s) for the theory.
Therefore, it is reasonable to ask how global parameters would be reflected in the Lagrangians of QT and QFT. 
The authors finding on this question is that, if (1) the causal model contains compound objects such as quantum objects which are not merely aggregations of smaller objects (for which the existence of global parameters is an indication) and (2) it is reasonable to represent the related laws of physics in the Lagrangian of the theory, the compound objects and the
related global parameters also have to be represented explicitly in the Lagrangian. 
In general, it may not be possible to include the compound, i.e., higher level objects in the Lagrangian that corresponds to the lower level entities. Therefore, multiple Lagrangians related to different levels of view may be required.

\section{Interactions between Quantum Objects}

For two reasons, the interactions between quantum objects are key for the local causal model of QT/QFT:
\begin{enumerate}
\item The causal model of the problem areas described in Section 2.2 (the measurement problem, the "interference collapse rule" and entanglement) is based
on the model of interactions between quantum objects.
\item Because quantum objects are assumed to run autonomously, their global relationships are mainly determined by the information exchange with interactions.
\end{enumerate}
In the causal model of QT/QFT, a quantum object  $ QO_{1}$  is considered to interact with another quantum object 
 $ QO_{2}$, if  $ QO_{1}$ shares some spacetime points with $ QO_{2}$.
In terms of wave equations (i.e., the equations of motion for the particles' waves)  (see \cite{Strassler}),  an interaction between two waves $ \psi_{1} $ and $ \psi_{2} $ resulting in a third wave $ \psi_{3} $ is described by an equation of motion in which the product of waves  $ \psi_{1} $ and $ \psi_{2} $ is related to $ \psi_{3} $, for example, in

 $ d^{2}\psi_{3}/dt^{2} - c^{2} d^{2}\psi_{3}/dx^{2} =  a^{2} \psi_{3} + b \cdot \psi_{1} \psi_{2} $. 
\\
Typical examples of interactions are particle scatterings such as, for example, electron-photon scattering. Interactions between quantum objects consisting of
particle collections (e.g., entangled particles or bound-system quantum objects) must be considered as well. 
\begin{figure}[ht]
\center{\includegraphics*[scale=0.8] {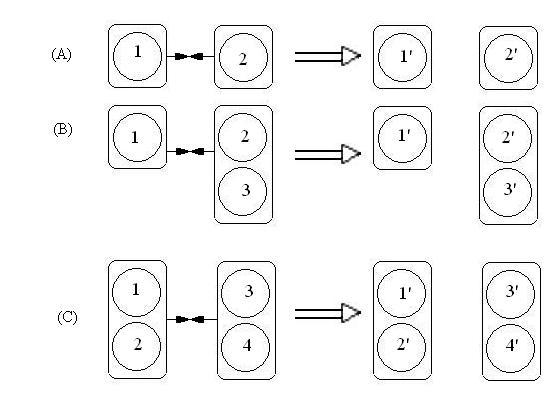} }
\caption{Examples of volatile interactions}
\end{figure}
In general, an interaction between two quantum objects may change the interacting quantum objects to differing extents. The changes may range from changed attributes (e.g., momenta and spins) to
changes in the numbers and types of the particles. Fig. 4 shows some simple types of interactions where it is assumed that only the attributes of the particles
change; the numbers and types of the outgoing particles are the same as the numbers and types of the ingoing particles. The interactions shown in Fig. 4 can be described by the
laws of (classical) quantum theory and are called "volatile interactions" in this paper.

In contrast to volatile interactions, there are also more complex interactions where the "out" quantum objects and/or the "out" particles (which are contained
in the "out" quantum objects) may differ from the "in" quantum objects and particles (see Fig. 5). Because such interactions require QFT (e.g., a scattering matrix
and Feynman diagrams) in their description, they are called "QFT interactions" within this paper. With QFT interactions, only a single path of the
"in" particles determines the interaction result.
\\
The overall causal model for the treatment of interactions is as follows:
\begin{verbatim}
Interaction-processing :=  {
overall-ialist = determine-potential-interactions();
IF not-empty( overall-ialist ) {
   QFT-ialist= determine-QFT-interactions(overall-ialist);
   IF is-empty( QFT-ialist ) {
      FOR ( all  interaction[i] from overall-ialist })  {
             perform-volatile-interaction( interaction[i] ); 
      }
   }
   ELSE  {
      //  perform QFT-interaction
      IF  size(QFT-ialist) > 1 {
           QFT-interaction = RANDOM(QFT-ialist );
      }
      ELSE {
          QFT-interaction = QFT-ialist[1];
      }
      perform-QFT-interaction(QFT-interaction); 
   }
}
}
\end{verbatim}
The functions determine-potential-interactions(), determine-QFT-interactions(),
perform-volatile-interaction( ) and perform-QFT-interaction( ) must be specified in more detail.

\subsubsection{determine-potential-interactions()} The potential interactions are identified by space points belonging to the processed particle; furthermore, they are shared by
another particle.
\\
 The proposed causal model of QT/QFT assumes that the physics engines of the individual quantum objects determine the occurrence of a (potential) interaction by checking if the space points occupied by the corresponding quantum object are also occupied by other quantum objects. This checking process is feasible only if
all of the (autonomous) quantum objects agree on a common global space coordinate system, or alternatively, if all of the space points have associated the quantum
objects and fields that occupy the space point.

\subsubsection{determine-QFT-interactions()} The criteria for the determination of QFT interactions are not quite clear to the author. The following criteria
influence the determination of QFT interactions:
\begin{itemize}
\item The particle types, i.e., whether QFT supports interactions between the particles.
\item The particles' probability amplitudes at the interaction position.
\item The particles' energy.
\end{itemize}
\subsubsection{perform-volatile-interaction( ) and perform-QFT-interaction()} are discussed below.

\subsection{Volatile Interactions}
Volatile interactions are interactions between particles known from classical quantum mechanics that change neither the type nor the number of the interacting particles. 
Volatile interactions initially can change only attributes which are related to the space-points of the affected path of the interacting particle. The space-point-related changes may propagate to the complete quantum object or to subunits (e.g., particles) belonging to the quantum object.
Fig. 4 shows some examples of volatile interactions. 
Because the laws of QT concerning volatile interactions are relatively straightforward, this type of interaction
is not discussed within this paper.

\subsection{QFT-Interactions}

With QFT interactions, the paths (i.e., space points) that triggered the interaction exclusively determine the outcome of the interaction. 
\footnote{ Therefore, QFT interactions are suited for QT measurements.}
\begin{figure}[ht]
\center{\includegraphics*[scale=0.6] {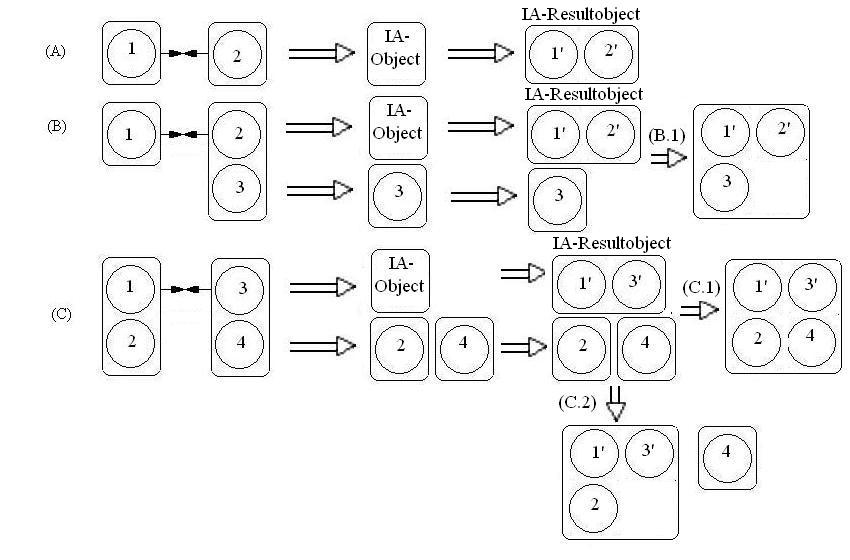} }
\caption{Examples of QFT-interactions}
\end{figure}
In the causal model of QT/QFT, the paths that do not participate in the interaction are discarded, and the interacting particles (not necessarily the interacting quantum
objects) are replaced by a single new quantum object called the interaction object ("IA-object” in Fig. 5). The destruction of the interacting particles and their
replacement by the interaction object may viewed as the collapse of the wave functions.

In contrast to volatile interactions (see above), with QFT interactions the numbers and types of the "out" particles may differ from the  "in" particles. The   
detailed laws of physics for the treatment of QFT interactions are given by QFT.
The overall causal model for the treatment of QFT interactions is as follows:
\begin{verbatim}
perform-QFT-interaction( interaction) := {
   IA-Object=create-interaction-object(interaction);
   drop-particle(  interaction.particle1);
   drop-particle(  interaction.particle2);
   FOR ( all other paths in quantumobject.path[i]) {
         eliminate-unaffected-paths(quantumobject.path[i]); 
   }
   ia-resultobject =  process-interaction-object(IA-Object);
}
\end{verbatim}
Fig. 5 shows some examples of QFT interactions: (A) the interaction between two single particles, (B) the interaction between a single particle and a
quantum object containing two particles, and (C) the interaction between two quantum objects each containing two particles. In each case, the two interacting
particles are first replaced by the interaction object ("IA-object" in Fig. 5). The processing of the interaction object is determined by the rules and equations of
QFT (e.g., Feynman diagrams, Feynman rules, and Fermion chains). However, these QFT rules must be mapped to a causal model. The details of this mapping are
described in 
 \cite{Dielfi} and   \cite{DielCALagr}. The overall result of the QFT interaction is embraced in a single particle collection (i.e., a quantum object), which we call
the ia-resultobject. The ia-resultobject typically contains two particles (which can be of the same type as the ingoing particles) and multiple paths.
The generation of a single particle collection ensures proper correlations with the alternative outcomes of the interaction. More details on 
process-interactionobject(IA-Object) are given in Section 5.2.

In cases where one of the "in" quantum objects consists of multiple particles, the "ia-resultobject" may further merge with the (remainder of the) "in"
quantum objects as indicated by (B.1), (C.1) and (C.2) in Fig. 5.

In cases where one or both of the interacting quantum objects contain more than a single particle (see Fig. 5, cases (B) and (C)), the question arises whether
the ia-resultobject may contain more particles in addition to the (replacement of) the directly interacting particles. The causal model of QT/QFT supports
this possibility by a continued processing following the creation of the ia-resultobject. In Fig. 5, this event is indicated by the steps labeled (B.1), (C.1) and (C.2).

\subsection{Summary on Interactions}

The major features of QFT interactions can be summarized as follows:
\begin{itemize}
\item A QFT interaction is caused by single definite space-time position (i.e., single paths of interacting quantum objects).
\item Among interacting quantum objects, at most a single QFT interaction can occur at a specific point in time. This point in time is locally determined.
Local simultaneity is well defined for the quantum object.
\item Although each interaction starts as a path-path interaction, it may evolve into an interaction involving larger scopes.
\item QFT interactions are always path-related and particle-related (at least initially).
\item Measurements typically require at least one QFT interaction between the measured quantum object and a quantum object belonging to the measurement
apparatus.
\end{itemize}
Table 3 summarizes the major interaction cases. 
\begin{table}
\caption{\label{label}Possible results of interactions between quantum objects}
\begin{tabular} { | c | c  | c | c | c | c | }
\hline
Interaction type & object-1 & object-2 & causal model actions   & examples  &  QT/QFT   \\

\hline

volatile   &  particle &  atom  & no path reduction,  &  quantum optics  &  QM  \\

 int., case (B)   &             &            &  no separation        &      &     \\
\hline

volatile  &  atom   &    atom    &  no path reduction,  & thermodynamics,  &  QM  \\

 int., case(C)    &             &               &  no separation        &  atom scattering   &        \\
\hline
QFT  &  particle      &     particle &  reduction,   & Double-slit,   &  QFT     \\

interaction  &            &                &  disentanglement,   &  Measurement,   &      \\

   case (A)      &                &    & interaction object  &  $ e^{-}+e^{+} \rightarrow  e^{-}+e^{+} $ &    \\

                         &                &    & (type change)  &   $ e^{-}+e^{+} \rightarrow  \tau^{-}+\tau^{+} $  &    \\
\hline
QFT,   &  particle   & atom,   & reduction,  &  particle absorption   & QFT   \\

interaction   &            &  entangled              &  disentanglement,   & by atom, measurement  &  ext.  \\

  case (B)    &       &  particles     & interaction object  & of entangled part.    &    \\
\hline
QFT,   &  atom   & atom   & reduction,  & scattering of   & QFT   \\

 interaction   &            &                &  disentanglement,   & atoms    &  ext.  \\

 case (C)     &                &    & interaction object  &    &    \\
\hline
\end{tabular}
\end{table} 
\\ Comments on Table 3:
\begin{itemize}
\item The interaction types "volatile, case(B)" and "volatile, case (C)" correspond to the examples shown in Fig. 4. 
The interaction types  "QFT, case(A)", "QFT, case(B)" and "QFT, case (C)" correspond to the examples shown in Fig. 5. 
\item "QM" means quantum mechanics. QFT techniques (e.g., Feynman diagrams)  are not required to specify the processing.
\item "QFT ext." stands for QFT extended and means that the existing standard QFT does not enable the derivation of a causal model for these types of
interactions.
\end{itemize}

\subsection{Major examples of causal models involving QFT-interactions}

\subsubsection{The Causal Model of Measurement}  is mainly based on the assumption that to obtain information about the measured quantum object, a QT measurement involves at least one QFT interaction between the measured quantum object and the measurement apparatus. All of the peculiarities reflected
in the QT measurement problem (e.g., the collapse of the wave function and the inability to measure certain observables concurrently) can be explained by the causal model
of QFT interactions
 (see \cite{Dielmeas}). 

\subsubsection{Causal Model of the EPR experiment} 
The EPR experiment (see \cite{EPR}), which is illustrated in Fig. 6, measures the spins of two entangled particles  (photons assumed in Fig. 6). Actual experiments such as \cite{Aspect} verified that the measurement result obtained with particle 1 influences the measurement result of particle 2. With the causal model of QT/QFT, this correlation of the measurement results can partly be explained by the assumption that the two photons belong to a common
quantum object with common paths for photon 1 and photon 2. 
The complete causal model of the EPR experiment includes a model of imperfect correlations
and assumes that the spin direction is a (quantum-object-) global parameter and the change of the global spin by the polarizers (see \cite{Dielmeas}).
\begin{figure}[ht]
\center{\includegraphics*[scale=1] {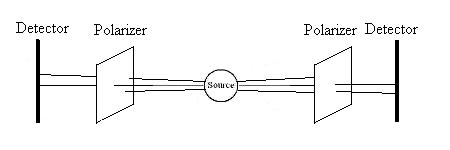} }
\caption{Components involved in the measurement of entangled photons}
\end{figure}

\subsubsection{The Causal Model of the Double-slit Experiment,} which is similar to the causal model of measurement, assumes that the collapse of the interference, which occurs with
special variants of the experiment, results from a QFT interaction occurring within one of the paths of the electron 
(see  \cite{Dieldslit}).

\section{Quantum Object Internal Dynamics}

A quantum object's internal dynamics can only partly be described in the form of a local causal model with precisely defined state transitions, a uniform state update
frequency and a locality restricted to a space-point locality. 
\footnote{ This fact supports the view of quantum objects as the elementary units of causality and locality.}
Nevertheless, it is possible to predict the final states of processes (such as particle
scattering), and it should be possible to identify certain intermediate states associated to sub-processes. As the first level of subdivision, the proposed local causal
model distinguishes a number of phases that a quantum object may pass or remain in:
\begin{enumerate}
\item Creation of the quantum object
\item Exchange of virtual particles ("Feynman phase") 
\item Entangled real particles ("EPR phase")
\item Non-entangled real particles  ("Heisenberg-Schr\"odinger phase")
\end{enumerate}
Depending on the type of quantum object, only part of the above list of phases is applicable. For the most general case, i.e., the interaction object, the complete list of
phases applies (see Section 5.2). The next important type of quantum object is the bound-system quantum object, where mainly the phase "Exchange of virtual
particles (Feynman phase)" is of interest.

\subsection{Phases of the  Quantum Object Internal Dynamics}

\subsubsection{Creation of a quantum object}
Quantum objects are created, transformed and extended as the result of interactions between quantum objects or decays of
quantum objects. The creation of the interaction object is the most important case of a quantum object's creation. With the creation of the interaction object, the
information from the two interacting particles is merged into the new interaction object.

\subsubsection{Feynman phase}
After a quantum object that consists of multiple particles (such as an interaction object) is created, the quantum object's internal dynamics
start with processes that are described by the laws of QFT. However, with QFT the rules that govern these processes are defined in terms of (external
and internal) lines and vertices of Feynman diagrams. For the causal model of QT/QFT, the QFT rules have to be mapped to rules regarding the components and
attributes of the interaction object. As described in Section 2.3, the causal model of QT/QFT is formulated largely in terms of discrete system state parameters
(space, time, and paths). This formulation enables the utilization (and sometimes requires the application) of some of the findings of lattice gauge theory (see, for example,
\cite{Kogut}). 
In standard QFT, bound systems (such as collections of quarks) can best be handled by the application of the concepts of lattice gauge theory. 

In \cite{Dielfi}, a more detailed description of a causal model of QFT interactions
is given with a focus on the Feynman phase. 
\footnote{ However, the name "Feynman phase" is not used in 
 \cite{Dielfi}.}

\subsubsection{EPR phase}
During the Feynman phase, the distances between the (virtual) particles that compose the interaction object are assumed to be sufficiently short 
that permanent interactions between the (virtual) particles occur. This phase (the Feynman phase) ends when a portion of the particles leaves the
scope of the local interactions. 
\footnote{Apparently, whereas interaction objects stay in the Feynman phase briefly, bound-system quantum objects have a longer lifetime.}
After leaving the Feynman phase, the causal model of QT/QFT assumes that the collection of particles remains a quantum
object, which means that the particles continue to be entangled through common alternative paths.

\subsubsection{Heisenberg-Schr\"odinger phase}
Here, the classical behavior of a non-entangled collection of particles is called the "Heisenberg-Schrödinger phase". Termination
of the entanglement occurs when a particle that is part of a quantum object interacts with another quantum object via a QFT interaction. The termination of the
entanglement typically implies the termination of the original quantum object. The termination of an existing quantum object always results in the formation of at least one new
quantum object. In Section 4, various cases of interactions between quantum objects with different resulting new quantum objects are discussed.

\subsection{Interaction Processing}

The processing of QFT-interactions (see Section 4.2) includes the following three phases:
\begin{enumerate}
\item Creation of the interaction object,
\item Processing of the interaction object (Feynman phase), 
\item Interaction result object (EPR phase).
\end{enumerate}
The major phase is the processing of the interaction object (the Feynman phase). The transition from the Feynman phase to the EPR phase is a statistical process
that could possibly be better understood by studying the discreteness of the causal model and its relationship to lattice gauge theory.

\subsection{Bound system's internal  dynamics}

The evolution of the bound-system quantum object is mainly within the Feynman phase, i.e., internal interactions between adjacent (virtual) particles belonging to
the bound-system quantum object. The Feynman phase may be (temporarily or permanently) interrupted by interactions with other quantum objects or by the
decay of the bound-system quantum object. In standard QFT, bound systems (such as collections of quarks) can best be handled by the application of the
concepts of lattice gauge theory. 

\section{Space-Time Considerations}

Although the integration of space and time as introduced with the theories of relativity remains valid in general, in the causal model of QT/QFT there are
situations where the treatment of time differs from that of space. As the major consequence of the autonomy of quantum objects, the time structure
is composed of local time units based on the quantum object's proper time. In contrast, space serves as the global medium for all inter-object relationships.

\subsection{Time Considerations}

In the proposed causal model, the progression of time is determined by the update frequency of the physics engine of the quantum object. Because each quantum
object is associated with its own physics engine, the quantum object is the smallest unit of simultaneity. Inertial systems composed of quantum objects with
equal velocity are compound units of simultaneity. The proper time clock rate (i.e., the update frequency of the physics engine) of the quantum object is
initially set when the quantum object is created as the result of a QFT interaction or a decay. The proper time interval is only modified due to interactions
with other quantum objects or due to interactions with fields. The quantum object's proper time determines
both the speed of all object-local state changes and the speed of the position changes within space. By determining the speed of the state changes and the speed of the position
changes, the proper time interval in conjunction with the mass of the quantum object determines completely the quantum object's energy.

\emph{physics engine clock rate = 1/proper time interval $ \propto $ energy } 

\subsection{Space Considerations}

Under the assumption that quantum objects are autonomous, a global medium must support the implementation of global physical processes and
object interrelationships. In physics in general, this role is given to space. To support this role, the quantum objects must refer to space-related
parameters in terms of a globally agreed addressing scheme. This setup includes references to spatial positions (i.e., the coordinates), the speed of the spatial position
changes (i.e., velocities), and directions in space.

In the proposed causal model, the following assumptions are made concerning
space:
\begin{itemize} 
\item Space is an active object, i.e., its evolution is driven by a physics engine with a global proper time interval.
\item Space is a superposition of all fields, i.e., fields are mapped to space and are updated by the global physics engine of the space.
\item A field expansion may imply spatial expansion or may occur within the existing space.
\item Mathematically, the space may be represented by a manifold (including the curvature).
\item Space changes dynamically, and it expands according to the expansions of the embedded fields. The spatial structure changes as the result of changes
(e.g., position changes) of the embedded quantum objects. As required by General Relativity, the energy (including masses) distribution of the clusters
of quantum objects determines the structure of the space.
\item Space does not shrink or disappear (even if the originating fields disappear).
\end{itemize} 


\section{Discussion} 

\subsection{What it means to say a "quantum object is autonomous"}

In terms of the causal model, the phrase "a quantum object is autonomous" means that each quantum object is driven by its individual, private physics engine. Thus, each quantum object has its individual proper time clock rate, and the laws of physics (i.e., the laws of QT and QFT) that determine the
quantum object's dynamics must 
as much as possible depend on quantum-object-local system state only.

The autonomy of the quantum object ends where the global relationships between quantum objects and the relationships between quantum objects and space
(with the associated fields) are considered:
\begin{itemize}
\item The space, which is the single global-system state component, may affect the dynamics of the quantum object.
\item The interactions between quantum objects and the interactions between quantum objects and fields may terminate a quantum object, change the state of a
quantum object or result in the creation of new quantum objects.
\end{itemize}

\subsection{Does a Black Hole represent a (single) quantum object?}

If a black hole has an internal structure and internal dynamics, it appears obvious that the laws of QT must apply to these internal relationships.
Therefore, a black hole should be viewed as a quantum object.

If a black hole is considered to be a quantum object, then it would adhere to the space-time concept described above (Section
6). Consequently, a black hole represents not a space-time singularity, but an elementary unit of space-time (in common with
quantum objects in general). In addition, new thoughts on the subject of black hole evaporation may be appropriate.
It is an open question whether a black hole can have multiple paths and entangled particles.

If the black hole is considered to be a quantum object, it would be a kind of bound system quantum object. As described in Section 5.3, bound system quantum objects remain in the Feynman phase.

\subsection{Does a field represent a quantum object?}

In the literature on QT, particles are frequently called quanta of fields  (see \cite{Strassler}). Fields have much in common with particles and particle  collections (i.e., quantum objects). Therefore, we must ask whether it is reasonable to consider a field as a type of quantum object. For the causal model of QT/QFT and the
quantum-object model described in this paper, it has been decided not to extend the definition of a quantum object to include fields, but rather to consider fields
as state components that are associated to space (see Section 6.2).

\section{Conclusions}

The authors attempt to construct a local causal model of QT/QFT  resulted in the identification of quantum objects as suitable elementary objects for this type of model. The quantum object is not merely a suitable construct for the description of the proposed model. Instead, the assumption
of quantum objects has implications that enable new concepts and solutions in a number of (problem) areas in QT/QFT, which is especially true if quantum
objects are assumed to be autonomous entities (see Section 7.1). For the model definition (and description),
the assumption of autonomous quantum objects requires strict separation of (a) quantum objects' internal state components, attributes and processes and (b)
inter-quantum-object physical relationships; the latter are represented in interactions between quantum objects and interactions between quantum objects and
space.

The introduction of quantum objects has also been motivated by the finding that a local causal model of QT/QFT is not feasible if a strong interpretation of
causality (i.e., precisely defined state transitions and a uniform state update frequency) and locality (i.e., space-point locality) is assumed. However, a local causal
model of QT/QFT is achievable if a weaker interpretation of causality and locality is tolerated within quantum objects. This possibility justifies the view
of quantum objects as elementary units of locality and causality.

%

\end{document}